\documentclass[showpacs,superscriptaddress]{revtex4}
\usepackage{amssymb}
\usepackage{amsmath}
\usepackage{graphicx}
\usepackage{subfigure}
\usepackage{natbib}
\usepackage{epsfig}
\usepackage{amsfonts}
\usepackage{mathrsfs}
\usepackage{CJK}
\usepackage[toc,page,title,titletoc,header]{appendix}

\begin{document}
\title{Enhancing entanglement trapping by weak measurement and quantum measurement reversal}
\author{Ying-Jie Zhang }
\affiliation{Shandong Provincial Key Laboratory of Laser
Polarization and Information Technology, Department of Physics, Qufu
Normal University, Qufu 273165, China}
 \affiliation{Institute of
Physics, Chinese Academy of Sciences, Beijing, 100190, China}

\author{Wei Han}
\affiliation{Shandong Provincial Key Laboratory of Laser
Polarization and Information Technology, Department of Physics, Qufu
Normal University, Qufu 273165, China}

\author{Heng Fan }
\email{hfan@iphy.ac.cn}
 \affiliation{Institute of Physics,
Chinese Academy of Sciences, Beijing, 100190, China}
\author{Yun-Jie Xia }
\email{yjxia@mail.qfnu.edu.cn}
 \affiliation{Shandong Provincial Key
Laboratory of Laser Polarization and Information Technology,
Department of Physics, Qufu Normal University, Qufu 273165, China}

\date{\today}
\begin{abstract}
In this paper, we propose a scheme to enhance trapping of
entanglement of two qubits in the environment of a photonic band gap
material. Our entanglement trapping promotion scheme makes use of
combined weak measurements and quantum measurement reversals. The
optimal promotion of entanglement trapping can be acquired with a
reasonable finite success probability by adjusting measurement
strengths.

\end{abstract}

\maketitle

\section{Introduction}

Entanglement is a vital resource for quantum information processing
such as quantum computation, quantum metrology and quantum
communication \cite{Nielsen-Chuang}. However, realistic quantum
systems are never completely isolated from the environment. The
inevitable interaction between a system and its environment leads to
quantum decoherence \cite{Zurek}. For an open multipartite quantum
system, decoherence leads to degradation of entanglement and, for
some cases, entanglement sudden death (ESD)
\cite{Yu,Almeida,Laurat,Xu,Zhang,ParaoanuGS}. Thus, tackling
decoherence for entanglement protection is a critical issue for
quantum information processing. It is therefore of interest to
examine the possible schemes that can lead to promotion or
preservation of entanglement.

At present, many methods have been proposed to protect entanglement
from decoherence and to increase the entanglement such as by
entanglement distillation \cite{Bennett,Pan,Dong}. Quantum Zeno
effect \cite{Facchi} can also be used to manipulated the decoherence
process, but in this method some special measurements should be
performed very frequently to freeze the quantum state in order to
prevent the degradation of entanglement. We can also deal with
decoherence by introducing the decoherence-free subspace
\cite{Lidar,Kwiat}. However, the decoherence-free subspace requires
the interaction Hamiltonian to have an appropriate symmetry, which
might not always be present. In most cases, the energy dissipation
of individual subsystems of a composite system is responsible for
the entanglement degradation. Hence, methods that can prevent the
decay of the excited-state population would be applicable. One way
widely applied is to place the qubits in a structured environment,
say, microcavity \cite{Osnaghi,Hagley} or in the photonic band gap
of photonic crystals \cite{Konopka,Zhangxia,LoFrancoB}. {\it{In
particular, in the photonic band gaps so as to inhibit spontaneous
emission, a trapping state is formed and permanent entanglement is
observed. This phenomenon, known as "entanglement trapping"
\cite{LoFranco,Lazarou,Bellomo}, can lead to effective long-time
entanglement protection.}}

 Recently, it is shown that weak
measurement and quantum measurement reversal can effectively
suppress amplitude-damping decoherence for a single qubit
\cite{Korotkov,Paraoanu,Lee}. For the case of two qubits,
remarkably, the weak measurement and quantum measurement reversal
can increase the entanglement, and even can avoid entanglement
sudden death \cite{Kimys}, see also \cite{ManXiaAn}. {\it{For weak
measurements \cite{Korotkov1}, the outcome cannot determine the
state of the measured system precisely and therefore does not
totally collapse the state of the system. The correspondingly
partial information is drawn from the measurement yielding a
nonunitary, nonprojective transformation of the quantum state.
Measurement reversal \cite{Katz1,Korotkov2} is a probabilistic
reversal of a partial quantum measurement, and only certain outcomes
of the measurement keep the full information of the initial state
and are possible to reverse. The probability of success decreases
with increasing strength of measurement, so that the reversible
measurement has zero probability for a traditional projective
measurement. Probabilistic reversal with a weak measurement has
already been demonstrated on a superconducting phase qubit
\cite{Katz1}, as well as on a photonic qubit \cite{KimYH}.}}

Then, it will be interesting to know whether the method of weak
measurement and quantum measurement reversal can be applied to
enhance the entanglement trapping in a common photonic band gap. In
this article, we show that this method indeed works for this system,
and in particular, the entanglement can be trapped in a higher
level. The success of this scheme is based on the fact that weak
measurement can be reversed and thus the amplitude-damping, the main
decoherence in photonic band gap, can be suppressed. We remark that
this scheme does not need frequent measurements compared with
quantum Zeno effect in suppressing decoherence.

This paper is organized as follows. In Sec. II, we describe
the model of two qubits interacting with environment of a photonic band gap. We
adopt the pseudomode approach to derive their evolution process. In
Sec. III, we propose the scheme to enhance the entanglement
trapping by using weak measurement and quantum measurement reversal.
Finally, in Sec. IV, we present the feasibility of the experimental
implementation of this scheme, and provide a brief conclusion.

\section{Physical model and dynamics process}

We consider a two-qubit system interacting with a common
zero-temperature bosonic reservoir. Our chosen specific system
consists of two identical two-level atoms ($A$ and $B$) interacting
with a common photonic band gap. The dynamics of two qubits coupled
to the reservoir modes can be describe by the Hamiltonian
\begin{eqnarray}
H=\omega_{0}\sigma^{A}_{+}\sigma^{A}_{-}+\omega_{0}\sigma^{B}_{+}\sigma^{B}_{-}+\sum_{k}\omega_{k}a^{\dag}_{k}a_{k}+[(\sigma^{A}_{+}+\sigma^{B}_{+})\sum_{k}g_{k}a_{k}+h.c.],\label{01}
\end{eqnarray}
where $a^{\dag}_{k}$, $a_{k}$ are the creation and annihilation
operators of quanta of the reservoir,
$\sigma^{j}_{+}=|e_{j}{\rangle}{\langle}g_{j}|$,
$\sigma^{j}_{-}=|g_{j}{\rangle}{\langle}e_{j}|$ and $\omega_{0}$ are
the inversion operators and transition frequency of the $j$-th qubit
(j=$A$, $B$); $\omega_{k}$ and $g_{k}$ are the frequency of the mode
$k$ of the reservoir and its coupling strength with two qubits.

 In order to find the dynamics of two qubits, {\it{we
solve the master equation by using the pseudomode approach
\cite{Garraway,Garraway1,Garraway2}.}} This exact master equation
describes the coherent interaction between the qubits and the
pseudomodes in presence of decay of the pseudomodes due to the
interaction with a Markovian reservoir. The number of the
pseudomodes relies on the shape of the reservoir spectral
distribution. We focus on an idealized model \cite{Garraway} of a
photonic band gap (or photon density of states gap) in which both
Lorentzians are centered at the same frequency, and one of them is
given a negative weighting, so that
$D(\omega)=\frac{W_{1}\Gamma_{1}}{(\omega-\omega_{c})^{2}+(\Gamma_{1}/2)^{2}}-\frac{W_{2}\Gamma_{2}}{(\omega-\omega_{c})^{2}+(\Gamma_{2}/2)^{2}}$,
where the weights of the two Lorentzians are such that
$W_{1}-W_{2}=1$. The effect of the Lorentzian with negative weight
is to introduced a dip into the density of states function
$D(\omega)$ where the coupling of the qubit will be inhibited.
$\omega_{c}$ is the center of the spectrum, and $\Gamma_{1}$,
$\Gamma_{2}$ are the full widths at half maximum of two Lorentzians,
respectively. There are two poles in $D$ which are located at
$\omega_{c}-i\Gamma_{1}/2$ and $\omega_{c}-i\Gamma_{2}/2$, and there
is a change in sign of the residues of $D$ between these poles. So
the photonic band gap exists two pseudomodes $a_{1}$ and $a_{2}$
decaying with decay rates
$\Gamma'_{1}=W_{1}\Gamma_{2}-W_{2}\Gamma_{1}$ and
$\Gamma'_{2}=W_{1}\Gamma_{1}-W_{2}\Gamma_{2}$ respectively. {\it{Two
qubits do not couple to the first pseudomode $a_{1}$ at all, they
only interacts with the second pseudomode $a_{2}$ (the strength of
the coupling $\Omega$) which is coupled to the first one (the
strength of the coupling
$V=\sqrt{W_{1}W_{2}}(\Gamma_{1}-\Gamma_{2})/2$), and both
pseudomodes are leaking into independent Markovian environments. The
exact pseudomode master equation associated with the band-gap model
is given by
\begin{eqnarray}
\frac{d\rho}{dt}=&-&i[H_{eff},\rho]-\frac{\Gamma'_{1}}{2}(a^{\dag}_{1}a_{1}\rho-2a_{1}{\rho}a^{\dag}_{1}+{\rho}a^{\dag}_{1}a_{1})-\frac{\Gamma'_{2}}{2}(a^{\dag}_{2}a_{2}\rho-2a_{2}{\rho}a^{\dag}_{2}+{\rho}a^{\dag}_{2}a_{2}),\label{001}
\end{eqnarray}
here, the effective Hamiltonian of the total system in the
pseudomode theory can be expressed
\begin{eqnarray}
H_{eff}&=&\omega_{0}\sigma^{A}_{+}\sigma^{A}_{-}+\omega_{0}\sigma^{B}_{+}\sigma^{B}_{-}+\omega_{c}a^{\dag}_{1}a_{1}+\omega_{c}a^{\dag}_{2}a_{2}+\Omega[a_{2}(\sigma^{A}_{+}+\sigma^{B}_{+})+h.c.]+V(a^{\dag}_{1}a_{2}+a_{1}a^{\dag}_{2}).
\label{002}
\end{eqnarray}
}}
 To illustrate the entanglement dynamics of two initially
entangled qubits, we consider that two qubits are initially in the
Bell-like state
$|\Psi\rangle_{AB}=\cos\theta|ee\rangle+\sin\theta|gg\rangle$, with
$\theta\in(0,\pi)$. Assuming $t=0$ the photonic band gap is in the
vacuum state $|\bar{0}\rangle_{E}=\prod_{k=1}^{N}|0_{k}\rangle$,
corresponding to the pseudomode theory, $|\bar{0}\rangle_{E}$ equals
to $|0_{1}0_{2}\rangle_{E}$, then the total state can be written as
$|\Psi(0)\rangle_{ABE}=|\Psi\rangle_{AB}\otimes|0_10_2\rangle_{E}$.
The total system contains at most two excitations. In this case the
dynamics of two qubits can be effectively described by a four-state
system in which three states are coupled to the cavity mode in a
ladder configuration, and one state is completely decoupled from the
other states. In the basis
$\{|0\rangle=|gg\rangle,|+\rangle=(|eg\rangle+|ge\rangle)/\sqrt{2},|-\rangle=(|eg\rangle-|ge\rangle)/\sqrt{2},|2\rangle=|ee\rangle\}$,
the effective Hamiltonian of the total system can be rewritten as
\begin{eqnarray}
H_{eff}=2\omega_{0}|2\rangle\langle2|+\omega_{0}|+\rangle\langle+|+\omega_{c}a^{\dag}_{1}a_{1}+\omega_{c}a^{\dag}_{2}a_{2}+\sqrt{2}\Omega(a^{\dag}_{2}|0\rangle\langle+|+a^{\dag}_{2}|+\rangle\langle2|+h.c.)
+V(a^{\dag}_{1}a_{2}+a_{1}a^{\dag}_{2}). \label{03}
\end{eqnarray}
{\it{From the Hamiltonian given by Eq.(\ref{03}), the subradiant
state $|-\rangle$ does not decay, and the super-radiant state
$|+\rangle$ is coupled to states $|0\rangle$ and  $|2\rangle$ via
the second pseudomode. The transitions
$|0\rangle\rightarrow|+\rangle$ and $|+\rangle\rightarrow|2\rangle$
have the same frequencies and are identically coupled with the
pseudomode $a_{2}$. As we all know,
$|0\rangle\otimes|0_10_2\rangle_{E}$ is invariant in the evolution
process, while $|2\rangle\otimes|0_10_2\rangle$ will decay. Then the
total system state $|\Psi\rangle_{AB}\otimes|0_10_2\rangle_{E}$
evolves to
\begin{eqnarray}
|\Psi(t)\rangle_{ABE}&=&\cos\theta(C_1(t)|20_10_2\rangle+C_2(t)|+0_11_2\rangle+C_3(t)|+1_10_2\rangle+C_4(t)|00_12_2\rangle+C_5(t)|01_11_2\rangle\nonumber\\
&+&C_6(t)|02_10_2\rangle)+\sin\theta|00_10_2\rangle],\label{04}
\end{eqnarray}
where these evolution coefficients satisfy
$\sum_{i=1}^{6}|C_i(t)|^{2}=1$. These evolution coefficients
$C_i(t)$ ($i=1,2,...,6$) can be found numerically by differential
equations, and the set of differential equations associated to the
pseudomode master equation (\ref{001}) is
\begin{eqnarray}
i\dot{C}_{1}(t)&=&2\omega_{0}C_{1}(t)+\sqrt{2}{\Omega}C_{2}(t),\nonumber\\
i\dot{C}_{2}(t)&=&\omega_{0}C_{2}(t)+(\omega_{c}-i\Gamma'_{2}/2)C_{2}(t)+\sqrt{2}{\Omega}C_{1}(t)+2{\Omega}C_{4}(t)+VC_{3}(t),\nonumber\\
i\dot{C}_{3}(t)&=&\omega_{0}C_{3}(t)+(\omega_{c}-i\Gamma'_{1}/2)C_{3}(t)+\sqrt{2}{\Omega}C_{5}(t)+VC_{2}(t),\nonumber\\
i\dot{C}_{4}(t)&=&2(\omega_{c}-i\Gamma'_{2}/2)C_{4}(t)+2{\Omega}C_{2}(t)+\sqrt{2}VC_{5}(t),\nonumber\\
i\dot{C}_{5}(t)&=&(\omega_{c}-i\Gamma'_{1}/2)C_{5}(t)+(\omega_{c}-i\Gamma'_{2}/2)C_{5}(t)+\sqrt{2}{\Omega}C_{3}(t)+\sqrt{2}VC_{4}(t)+\sqrt{2}VC_{6}(t),\nonumber\\
i\dot{C}_{6}(t)&=&2(\omega_{c}-i\Gamma'_{1}/2)C_{6}(t)+\sqrt{2}VC_{5}(t).
\label{003}
\end{eqnarray}
According to the above evolutionary dynamics process, we would show
a scheme to enhance entanglement of two qubits in a common photonic
band gap model by using the combined weak measurements and quantum
measurement reversals in the next section.}}

\section{Scheme for enhancing entanglement trapping}

With regard to the initial state $|\Psi(0)\rangle_{ABE}$ of the
whole system, it is interesting to find that the entanglement
between two qubits can be trapped after a certain time $t$ in the
photonic band gap without any measurements to qubits
\cite{Zhangxia,LoFrancoB,LoFranco,Lazarou,Bellomo}. And we know that
entanglement can be protected and increased by weak measurement and
quantum measurement reversal \cite{Kimys}. It is natural to consider
the question: is it possible to have a larger entanglement by using
weak measurement and quantum measurement reversal while still with
entanglement trapping occurring in the photonic band gap? We find a
positive answer to this question, and next will present our
entanglement trapping promotion scheme.

Firstly,  before the qubits undergo decoherence, we perform a weak
measurement on these two qubits respectively, which partially
collapses the state towards $|gg\rangle$. The two-qubit weak
measurement is a non-unitary quantum operation, and can be written
as
\begin{eqnarray}
M_{wk}(p_{A},p_{B})&=&\left(\begin{array}{cc}1&0\\0&\sqrt{1-p_{A}}\\
\end{array}\right)\otimes \left(\begin{array}{cc}1&0\\0&\sqrt{1-p_{B}}\\
\end{array}\right),\label{05}
\end{eqnarray}
 where $0{\leq}p_{A}\leq1$ and $0{\leq}p_{B}\leq1$ are the weak measurement
strengths. {\it{We mainly focus on the condition that the same
measurements performing on two qubits $p_{A}=p_{B}=p$.}} The system
qubits after the weak measurement are less vulnerable to
decoherence, because of the computational basis state $|gg\rangle$
does not couple to the environment. So the initial state of two
qubits $|\Psi\rangle_{AB}$ becomes
\begin{eqnarray}
|\Psi(p,0)\rangle_{AB}&=&\frac{\cos\theta
(1-p)|ee\rangle+\sin\theta|gg\rangle}{\sqrt{\cos^2\theta(1-p)^2+\sin^2\theta}}.
 \label{06}
\end{eqnarray}

We let two qubits undergo a common photonic band gap
(decoherence quantum channel). By the calculation process in the
Sec. II, after some time of interaction between the system and the environment,
the total state $|\Psi(p,0)\rangle_{ABE}$ evolves to
\begin{eqnarray}
|\Psi(p,t)\rangle_{ABE}
&=&\frac{1}{\sqrt{P(p,t)}}[\cos\theta(1-p)(C_1(t)|20_10_2\rangle+C_2(t)|+0_11_2\rangle+C_3(t)|+1_10_2\rangle\nonumber\\
&+&C_4(t)|00_12_2\rangle+C_5(t)|01_11_2\rangle+C_6(t)|02_10_2\rangle)+\sin\theta|00_10_2\rangle],
\label{07}
\end{eqnarray}
with
$P(p,t)=\cos^2\theta(1-p)^2(|C_1(t)|^2+|C_2(t)|^2+|C_3(t)|^2+|C_4(t)|^2+|C_5(t)|^2+|C_6(t)|^2)+\sin^2\theta$.
The reduced density matrix $\rho_{AB}(p,t)$ of qubits can be
obtained from Eq.(\ref{07}) by tracing over the pseudomode degrees
of freedom
\begin{eqnarray}
\rho^{wk}_{AB}(p,t)=\frac{1}{P(p,t)}[a(p,t)|0\rangle_{AB}\langle0|+b(p,t)|2\rangle_{AB}\langle2|+c(p,t)|+\rangle_{AB}\langle+|+d(p,t)|2\rangle_{AB}\langle0|
+d^*(p,t)|0\rangle_{AB}\langle2|],\label{08}
\end{eqnarray}
with
\begin{eqnarray}
a(p,t)&=&\sin^2\theta+\cos^2\theta(1-p)^2(|C_4|^2+|C_5|^2+|C_6^2|),\nonumber\\
b(p,t)&=&\cos^2\theta(1-p)^2|C_1|^2,\nonumber\\
c(p,t)&=&\cos^2\theta(1-p)^2(|C_2|^2+|C_3|^2),\nonumber\\
d(p,t)&=&\cos\theta\sin\theta(1-p)C_1.\label{09}
\end{eqnarray}

In order to gain as much entanglement as possible,
we should perform the post-measurement on these two qubits after the system undergoing the
evolution process. As discussed in Ref. \cite{ManXiaAn}, we must
determine which post-measurement should be taken,
weak measurement or quantum measurement reversal,
by comparing the
value of $a(p,t)$ and $b(p,t)$. When $a(p,t)<b(p,t)$, the weak
measurement should be chosen as the post-measurement during the entangled qubits undergone
decoherence process. The operation is the same as Eq.(\ref{05}), but
the post-measurement (weak measurement) strengths are
replaced by $p_{r_A}=p_{r_B}=p_r\in[0,1]$. Then the evolutional
state becomes
\begin{eqnarray}
\rho^{r(1)}_{AB}(p,p_{r},t)&=&\frac{1}{P_1(p,p_{r},t)}[a(p,t)|0\rangle_{AB}\langle0|
+b(p,t)(1-p_r)^2|2\rangle_{AB}\langle2|+c(p,t)(1-p_r)|+\rangle_{AB}\langle+|\nonumber\\
&+&d(p,t)(1-p_r)|2\rangle_{AB}\langle0|+d^*(p,t)(1-p_r)|0\rangle_{AB}\langle2|],\label{10}
\end{eqnarray}
where $ P_1(p,p_r,t)=a(p,t)+b(p,t)(1-p_r)^2+c(p,t)(1-p_r)$ is the
overall success probability of the combined former and latter weak
measurements. On the other hand, when $a(p,t)>b(p,t)$, a non-unitary
quantum measurement reversal operation $M_{r}$ as the
post-measurement can be given to two qubits respectively during they
interact with their common photonic band gap, which is
\begin{eqnarray}
M_{r}(p_{r_{A}},p_{r_{B}})&=&\left(\begin{array}{cc}\sqrt{1-p_{r_A}}&0\\0&1\\
\end{array}\right)\otimes \left(\begin{array}{cc}\sqrt{1-p_{r_B}}&0\\0&1\\
\end{array}\right),
\label{11}
\end{eqnarray}
where $p_{r_A}$ and $p_{r_B}$ are the reversal measurement strengths
(here, $p_{r_A}=p_{r_B}=p_r\in[0,1]$). Under such
condition, we obtain
\begin{eqnarray}
\rho^{r(2)}_{AB}(p,p_{r},t)&=&\frac{1}{P_2(p,p_{r},t)}[a(p,t)(1-p_r)^2|0\rangle_{AB}\langle0|+b(p,t)|2\rangle_{AB}\langle2|+c(p,t)(1-p_r)|+\rangle_{AB}\langle+|\nonumber\\
&+&d(p,t)(1-p_r)|2\rangle_{AB}\langle0|+d^*(p,t)(1-p_r)|0\rangle_{AB}\langle2|],\label{12}
\end{eqnarray}
and $ P_2(p,p_r,t)=a(p,t)(1-p_r)^2+b(p,t)+c(p,t)(1-p_r)$ is the
success probability of prior weak measurements and the following
quantum measurement reversals.

 To quantify the entanglement, we use the concurrence \cite{Wootters}, defined as
$C(t)=\max\{{0,\sqrt{\lambda_{1}}-\sqrt{\lambda_{2}}-\sqrt{\lambda_{3}}-\sqrt{\lambda_{4}}}\}$,
where $\{{\lambda_{i}}\}$ are the eigenvalues of the matrix
$M=\rho(\sigma^{A}_{y}\otimes\sigma^{B}_{y})\rho^{*}(\sigma^{A}_{y}\otimes\sigma^{B}_{y})$
in decreasing order, with $\rho^{*}$ denoting the complex conjugate
of $\rho$, $\sigma^{A}_{y}$ and $\sigma^{B}_{y}$ are the Pauli
matrices for qubits $A$ and $B$. {\it{Here, we mainly examine the
entanglement of a class of important bipartite density matrices. A
density matrix in the class only contains non-zero elements along
the main diagonal and anti-diagonal, and is defined as $X$ state
\cite{YuTing,Heydari1},
\begin{equation}
\rho_{AB}=\left(
       \begin{array}{cccc}
         x & 0 & 0 & v \\
         0 & y & u & 0 \\
         0 & u^{\ast} & z & 0 \\
         v^{\ast} & 0 & 0 & w \\
       \end{array}
     \right),\label{004}
\end{equation}
with $x,y,z,w$ real positive and $u,v$ complex quantities. This $X$
mixed state arises naturally in a wide variety of physical
situations. Such as the Bell-like states as well as the well-known
Werner mixed state are classified to this form. Unitary transforms
of the $X$ state extend its domain even more widely. And the $X$
states defined above not only are rather common but also have the
property that they would retain the X form under decoherence
evolution. For the X state defined in Eq. (\ref{004}), concurrence
\cite{Wootters} can be simplified as
$C(\rho_{AB})=2Max\{0,|u|-\sqrt{xw},|v|-\sqrt{yz}\}$.

In this paper, the initial state of two qubits has an $X$ form, so
the two-qubits reduced density matrices
$\rho^{r(i)}_{AB}(p,p_{r},t)$ (with $i=1,2$) preserve the $X$ form
during the system evolution in the standard basis $\{|gg\rangle,
|ge\rangle, |eg\rangle, |ee\rangle\}$. So concurrence of
entanglement can be formally derived as
\begin{eqnarray}
C^i_{AB}(p,p_r,t)=\frac{2}{P_i(p,p_r,t)}Max\{0,
(1-p_r)(\frac{|c(p,t)|}{2}-\sqrt{a(p,t)b(p,t)}),
(1-p_r)(|d(p,t)|-\frac{|c(p,t)|}{2})\},\label{0013}
\end{eqnarray}
in fact, $(1-p_r)(\frac{|c(p,t)|}{2}-\sqrt{a(p,t)b(p,t)})$ is always
negative, then we eventually obtain}}
\begin{eqnarray}
C^i_{AB}(p,p_r,t)=\frac{2}{P_i(p,p_r,t)}Max\{0,
(1-p_r)(|d(p,t)|-\frac{|c(p,t)|}{2})\}.\label{13}
\end{eqnarray}

 Here the photonic band gap is acting as the two qubits
decoherence quantum channel. For a perfect gap, where
$D(\omega_{c})=0$ which can be satisfied if
$W_{1}/\Gamma_{1}=W_{2}/\Gamma_{2}$, there appear two-qubit
entanglement trapping if the qubits are resonant with the gap in the
weak-coupling regime \cite{Zhangxia,Lazarou}. In this paper, our
scheme is to make use of the pre-measurement (weak measurement) and
the post-measurement (weak measurement or quantum measurement
reversal) to enhance the two-qubit entanglement trapping in a common
photonic band gap.
\begin{figure}[tbh]
\includegraphics[bb=140 70 560 360, width=12 cm, clip]{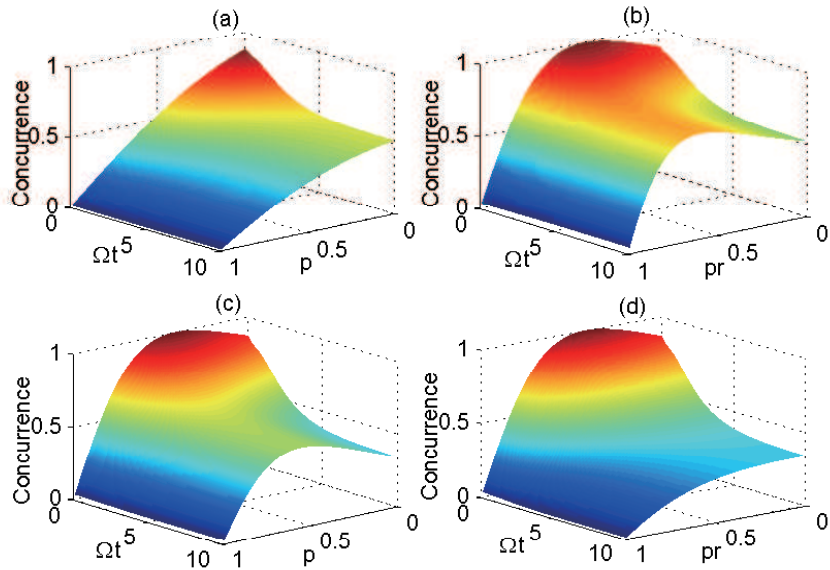}
\caption{(Color online) The two-qubit
concurrence as functions of the dimensionless quantity ${\Omega}t$
and measurement strength $p$ (or $p_r$) in a common photonic band
gap. The parameters used here: $W_{1}=1.1$, $W_{2}=0.1$,
$\Gamma_{1}=11\Omega$, $\Gamma_{2}=\Omega$ satisfy the weak-coupling
regime. For the cases (a)(b) $\theta=\pi/3$, (c)(d) $\theta=\pi/6$.
And (a)(c) $p_r=0$, i.e., only a pre weak measurement performed on
the initial state, (b)(d) no pre weak measurement $p=0$, only making
post-measurement (weak measurement or quantum measurement reversal)
to two qubits, respectively.}
\end{figure}

\begin{figure}[tbh]
\includegraphics[bb=50 185 425 300, width=12cm, clip]{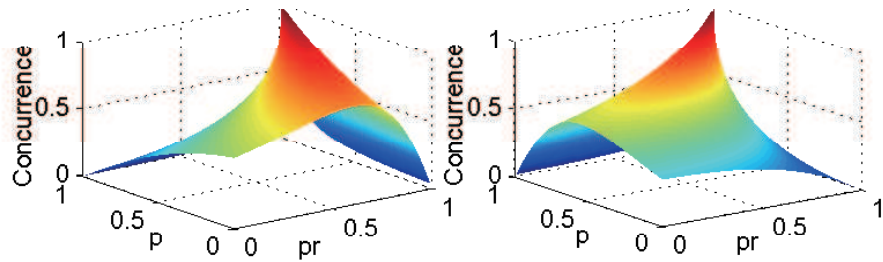}
 \caption{(Color online) The value of
two-qubit entanglement trapping as functions of the pre-measurement
strength $p$ and post measurement strength $p_r$ when the
entanglement trapping occurs (here choosing $\Omega t=15$). The
parameters used are: $W_{1}=1.1$, $W_{2}=0.1$,
$\Gamma_{1}=11\Omega$, $\Gamma_{2}=\Omega$. For the cases (a)
$\theta=\pi/3$, (b) $\theta=\pi/6$.}
\end{figure}

In Fig. 1, we show respectively the effect of the weak measurement
or quantum measurement reversal on entanglement trapping of
different initial states ($\theta=\pi/3$ and $\theta=\pi/6$). We
demonstrate clearly that only taking the pre-measurements ($p\neq0$,
$p_r=0$) or the post-measurements ($p_{r}\neq0, p=0$) to these two
qubits can enhance two-qubit entanglement trapping. We note that,
the edge of each graph in Fig. 1 corresponding $p=0$ or $p_{r}=0$
represents the entanglement trapping obtained without the action of
any measurement. In the case $\theta=\pi/3$, the weight of
$|ee\rangle$ is less than that of $|gg\rangle$ in the initial state,
if we only carry out the pre-measurements on the qubits, the
two-qubit entanglement trapping cannot be improved by comparing with
the concurrence of entanglement trapping without introducing any
operations to qubits, as shown in Fig. 1(a). While in the absence of
the pre-measurement but only performing the post-measurements
(quantum measurement reversal) to these two qubits respectively, the
two-qubit entanglement trapping can be enhanced in a certain region
of $p_r$, as shown in Fig. 1(b). This is because the quantum
measurement reversals can decrease the $|gg\rangle$ component such
that enhancement of entanglement trapping can be achieved.

In contrast, when $\theta=\pi/6$, the weight of $|ee\rangle$ is more
than the weight of $|gg\rangle$, Figs. 1(c) and 1(d) reveal that
two-qubit entanglement trapping can be promoted in the case of only
the pre-measurements performed on the initial state within a
specific $p$ range. When only post-measurements are performed to two
qubits, the concurrence of entanglement trapping is smaller than
that obtained without doing any operations to qubits. That means
post-measurements are not necessary for this case. This is easy to
understand that the pre-measurements reduce the $|ee\rangle$
component to suppress decoherence process. So two-qubit entanglement
trapping can be promoted by using mainly the pre-measurements (weak
measurement) to qubits when the $|ee\rangle$ weight is larger than
the $|gg\rangle$ weight in the initial states. But if the
$|ee\rangle$ weight is smaller than the $|gg\rangle$ weight in the
initial states, quantum measurement reversals play a key role to
enhance the two-qubit entanglement trapping as we mentioned
previously.

Our aim is to promote the entanglement while keeping entanglement
trapping by making both the pre-measurements (weak measurement) and
the post-measurements (weak measurement or quantum measurement
reversal). Next, we study the roles of prior weak measurement
strength $p$ and post-measurement strength $p_r$ on concurrence of
two-qubit entanglement trapping. In Figs. 2(a) and 2(b), we show the
concurrence of entanglement trapping influenced by $p$ and $p_r$ for
$\theta=\pi/3$ and $\theta=\pi/6$, and the rescaled time $\Omega
t=15$ is fixed at which the phenomenon of two-qubit entanglement
trapping has already occurred. From Fig. 2(a), we have verified that
when the pre-measurement strength $p$ is given, the largest
concurrence of entanglement trapping can be acquired through an
optimal post-measurement strength $p_r$. And the largest concurrence
and its corresponding optimal post-measurement strength $p_r$ both
can increase with the pre weak measurement strength $p$ increasing.
By contrast, in the case $\theta=\pi/6$, as shown in Fig. 2(b), the
largest concurrence of entanglement trapping should relate closely
with an optimal pre-measurement strength $p$ when the
post-measurement strength $p_r$ is fixed. Moreover, the largest
entanglement trapping and its corresponding optimal pre-measurement
strength $p$ both can increase with the post-measurement strength
$p_{r}$ increasing.

\begin{figure}[tbh]
\includegraphics[bb=70 150 480 340, width=12cm, clip]{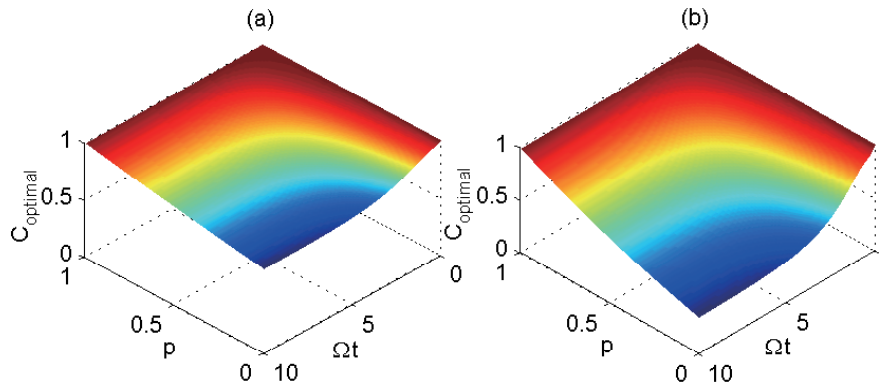}
 \caption{(Color online) The optimal
concurrence of two qubits as functions of the dimensionless quantity
${\Omega}t$ and the pre-measurement strength $p$. (a)
$\theta=\pi/3$, (b) $\theta=\pi/6$.}
\end{figure}

In view of the above analysis, we have two control parameters $p$
and $p_{r}$ to promote the qubits entanglement trapping. In order to
obtain the most effective entanglement trapping, we note that the
concurrence does not vary monotonically with $p$ for a fixed
$p_{r}$, and also not vary monotonically with $p_{r}$ for a fixed
$p$. Next, we investigate the optimal condition which maximizes the
concurrence of entanglement trapping. By solving the extreme value
of $C^i_{AB}(p,p_r,t)$ in Eq.(\ref{13}), the optimal
post-measurement strength $p_{r}$ that gives the maximum concurrence
of $\rho^{r(i)}_{AB}(p,p_{r},t)$ can be calculated as following: (i)
for the case of $a(p,t)<b(p,t)$, $p_{r}$ and $p$ satisfies
$p_{r}=1-\sqrt{a(p,t)/b(p,t)}$, and (ii) in the case
$a(p,t)>b(p,t)$, yielding $p_{r}=1-\sqrt{b(p,t)/a(p,t)}$. Then in
Figs. 3(a) and 3(b) we show the evolution of the largest concurrence
of entanglement trappings $C_{optimal}$ corresponding to the above
optimal conditions, depending on $p$ for $\theta=\pi/3$ and
$\theta=\pi/6$. We see that the optimal entanglement trapping
increase with increasing $p$ and in principle the qubits could be
trapped in the maximal entanglement state if $p$ is chosen close to
$1$. A comparison between Fig. 3(a) and Fig. 3(b) reveals that for
different initial entanglement states, two-qubit entanglement
trapping can get different promotion with an identical
pre-measurement strength $p$. When the entangled state has been
trapped (setting ${\Omega}t=15$), Fig. 4(a) shows the concurrence of
the optimal entanglement trapping as functions of $p$ by choosing
different initial states. It is clear that the concurrence of
optimal entanglement trapping, which corresponds to a same $p$,
increases with increasing $\theta$ in the region
$\theta\in(0,\pi/2)$. However, the corresponding success probability
$P_{optimal}$ decreases with the initial condition $\theta$
increasing, and also decreases monotonically with $p$ increasing, as
shown in Fig. 4(b). Thus by fixing the pre-measurement strength $p$
not close to 1 for an initial entangled state, we may find the
optimal promotion of entanglement trapping by our scheme with
acceptable success probability $P_{optimal}$.

\begin{figure}[tbh]
\includegraphics[bb=12 10 210 125, width=12cm, clip]{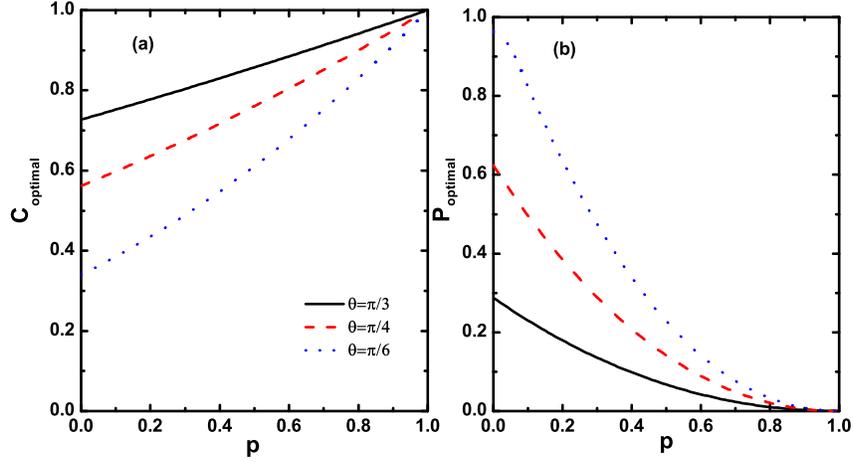}
 \caption{(Color online) The optimal concurrences $(a)$ and the
corresponding success probabilities $(b)$ under our promotion scheme
as functions of $p$ at different initial states when the
entanglement trapping has occurred (here considering
${\Omega}t=15$). The parameters used are: $W_{1}=1.1$, $W_{2}=0.1$,
$\Gamma_{1}=11\Omega$, $\Gamma_{2}=\Omega$.}
\end{figure}

\begin{figure}[tbh]
\includegraphics[bb=20 130 480 310, width=12cm, clip]{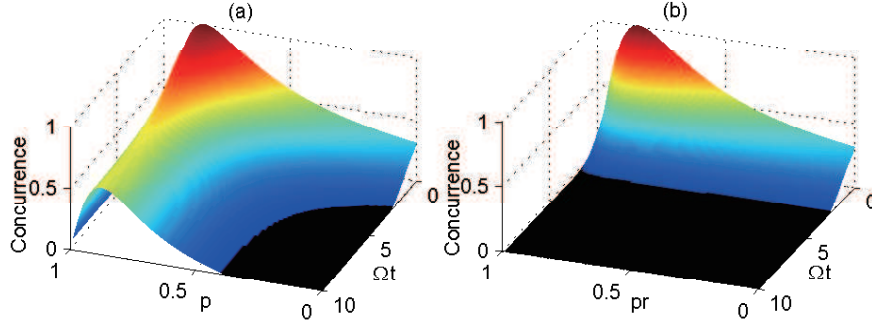}
 \caption{(Color online) The two-qubit
concurrence as functions of the dimensionless quantity ${\Omega}t$
and the measurement strength $p$ (or $p_r$) in a common photonic
band gap. The parameters used are: $\theta=\pi/20$, $W_{1}=1.1$,
$W_{2}=0.1$, $\Gamma_{1}=11\Omega$, $\Gamma_{2}=\Omega$. For the
cases (a) $p_r=0$, (b)$ p=0$.}
\end{figure}

 Finally, it is worth noting that ESD can also occur in the
photonic band gap material when the $|ee\rangle$ weight is much
larger than the $|gg\rangle$ weight. In Fig. 5, by choosing the
initial state $\theta=\pi/20$, ESD appears in the case $p=p_{r}=0$
(two qubits without any measurements ). We illustrate the
concurrence as functions of rescaled time $\Omega t$ and weak
measurement strength $p$, as shown in Fig. 5(a). It is clear to see
that ESD will never occur and entanglement trapping would appear
under the condition that weak measurement strength $p$ larger than a
certain value. As an example, for the parameters used in Fig. 5(a),
the concurrence will never vanish when $p>0.4$. To find out the
effect of the post measurement strength $p_{r}$ to ESD, we also
display the evolution of two-qubit concurrence depending on $p_{r}$
in the case $p=0$ in Fig. 5(b). As can be seen, in the absence of
the pre-measurements, performing the post measurements alone cannot
circumvent ESD-causing. So in order to restrain ESD-causing in a
photonic band gap, one should apply weak measurements with a quite
large $p$, as mentioned in Ref. \cite{Kimys} where two
amplitude-damping decoherence channels are considered.

\section{Discussion and conclusion}

The entanglement trapping promotion scheme presented in this paper
is valid for ideal photonic band gap. In real crystals with
finite dimensions, a pseudogap corresponding to the photonic band gap
can be obtained where the density of
states is much smaller than that of free space, though it is not
exactly zero. In these gaps, the spontaneous decay of an excited
emitter is inhibited by setting its suitable position  inside the
photonic band gap materials \cite{Sprik,Busch,Woldeyohannes}.
This system is similar to the ideal photonic band gap for
processing the scheme proposed in this paper.
Hence, entanglement trapping and promotion can occur in these materials.

In our scheme, two qubits must be entangled initially in structured
environments. In experiment, this initial entangled state can be
generated when a pair of atoms coupled near-resonantly to the edge
of photonic band gap have direct dipole-dipole interaction
\cite{Nikolopoulos}. Or alternatively, the entangled state for
spatially separated Rydberg-atoms can be realized by choosing a
three-dimensional photonic crystal single-mode cavity with
high-quality $Q$ factor where atoms can freely travel through the
connected void regions \cite{Guney}. And in the Rydberg-atom
context, by inserting a defect mode as a cavity inside the crystal,
suitable atom-cavity interactions allow one to perform quantum logic
gates and other cavity QED-based quantum state manipulations
\cite{Guney}. In other systems, for example, the coherent control of
an exciton in a quantum dot is also experimentally achievable
\cite{Bonadeo}. Concerning about the experimental feasibility of our
scheme, weak measurement and quantum measurement reversal operations
for a single qubit and two qubits have already been demonstrated
successfully \cite{Korotkov,Kimys,Katz1,KimYH}. {\it{On the basis of
the above analysis, our scheme might be implemented with current
experimental technologies. In fact, the experiments are still quite
challenging and there are a lot of subtle aspects to
implementations.}}

In conclusion, we have presented the promotion entanglement trapping
scheme by means of weak measurements and quantum measurement
reversals. In particular, for a photonic band gap as the decoherence
channel, we have shown that our protocol can enhance two-qubit
entanglement trapping. We have also analyzed relationships about the
optimal entanglement trapping, the corresponding success probability
and weak measurement strength. Moreover, we indicate that the
pre-measurement can be used to prevent ESD in the photonic band gap,
comparatively the post measurements alone cannot circumvent
ESD-causing. The evidences obtained show that entanglement trapping
can be effectively promoted by using weak measurements and quantum
measurement reversals. This highlights the potential of reservoir
engineering for controlling and manipulating the dynamics of quantum
systems.

\section{Acknowledgments}
This work is supported by National Natural Science Foundation of
China under Grant Nos. 11175248, 61178012, 11247240, 11304179, the
Specialized Research Fund for the Doctoral Program of Higher
Education under Grant Nos. 20123705120002, 20133705110001, the
Provincial Natural Science Foundation of Shandong under Grant No.
ZR2012FQ024, the Scientific Research Foundation of Qufu Normal
University for Doctors under Grant No. BSQD20110132.

\end{document}